\documentclass[12pt]{article}
\usepackage{amsmath}
\usepackage{amstext}
\usepackage{amsfonts}
\usepackage{amsbsy}
\usepackage{eucal}
\usepackage{amssymb}
\usepackage{amsthm}
\usepackage{color}
\usepackage{graphics}
\usepackage{color}
\usepackage[dvips]{graphicx}

\newcommand{\be}{\begin{equation}}
\newcommand{\ee}{\end{equation}}
\begin{document}
\begin{center}
{\bf Bilinear approach to Kuperschmidt super-KdV type equations}
\end{center}
\begin{center}
Corina N. Babalic$^{**}$, A. S. Carstea${^{*,**}}$
\end{center}
\begin{center}
{\it * Dept. of Theoretical Physics, Institute of Physics and Nuclear Engineering,  407 Atomistilor, Magurele, 077125 Bucharest, Romania\\
** University of Craiova, 13 A.I. Cuza, 200585, Craiova, Romania }
\end{center}
\begin{abstract}
Hirota bilinear form and soliton solutions for super-KdV of Kuperschmidt (Kuper-KdV) are given. It is shown that even though the collision of supersolitons is more complicated than in the case of 
supersymmetric KdV of Manin-Radul, the asymptotic effect of the interaction is simpler.
As a physical application it is shown that the well known FPU problem having a phonon-mediated interaction of some internal degrees of 
freedom expressed through grassmann fields, goes to the Kuper-KdV equation in a multiple-scale approach.
\end{abstract}

\section{Introduction}
Nonlinearly coupled equations containing bosonic and fermionic fields can display a very interesting phenomenology. The first systematic construction of an integrable supersymmetric hierarchy was given 
long time ago by Yu. Manin and A. Radul \cite{manin}. Complete integrability of such equations is a very interesting and deep problem which has no definitive answer so far 
both classically and quantum mechanically. However there are a lot of interesting results about integrability of {\it supersymmetric} nonlinear equations (supersymmetry being an extra symmetry imposed 
by construction which means, roughly speaking an invariance with respect to a kind of interchanging between the bosonic and fermionic fields)\cite{aratyn,manin}. Results about Darboux transformations \cite{liu1,kul1}, 
bi-hamiltonian structure \cite{fig}, 
prolongation structures \cite{pro} and Painlev\'e property \cite{mat} and many others revealed many nontrivial and completely 
new aspects about such systems. We also mention here appearance of non-local integrals of motion \cite{popo}, lack of unique bilinear formulation of integrable hierarchies \cite{kac1,kac2,leclair,ueno,yung}, 
inelastic interactions of super-solitons\cite{carstea1}, 
peculiar properties of super-Painlev\'e equations \cite{carstea4} etc.

From the point of view of solutions, the inelastic character of super-solitons interaction was observed in \cite{carstea0,carstea1} together with a dressing of fermionic phases. 
The study of super-soliton interaction was extended to lattice equations. Here the supersymmetry is broken due to 
discretization but still the equations can be analyzed using Lax pair or super-bilinear formalism \cite{levi,fane,grahov,mikh}

Case of non-supersymmetric coupled nonlinear equations is more difficult. Quite interesting the first completely super-extension of KdV was given independently by Kuperschmidt 
and Kulish \cite{kupers,kul2} before the 
Manin-Radul \cite{manin} paper. It is not supersymmetric and served as the first example of integrability in the grassmann algebra. Although the Lax pair and bi-hamiltonian structure were established 
quickly by Kuperschmidt himself, 
nothing was done about the solutions so far. The only partial answer was proposed recently in the paper of Kulish and Zeitlin \cite{kulish} where the IST scheme was adapted to Kuper-KdV 
equation but only in the case of a grassmann algebra with only one generator (case in which Kuper-KdV is reduced to a simple coupling between ordinary KdV and an ordinary linear equation).
Another non-supersymmetric system was the SSH-polyacetylene model which, in the non-resonant multiple scales, goes to purely fermionic complex-mKdV equation and, at the resonance between 
phononic and electronic dispersion branches, to coupled bosonic-fermionic Yajima-Oikawa-Redekopp equation having super-dark-solitons\cite{ssh1,ssh2}.

In this paper we are trying to go one step further and to compute super-soliton solution for Kuper-KdV equation using super-Hirota bilinear formalism. We will find a bilinear system 
containing some auxiliary fermionic tau function. We will examine the super-solitons interaction showing some similarities/differences with the Manin-Radul supersymmetric KdV case.

\section{Kuper-KdV type equations}
Throughout the paper we will deal with functions depending on $x, t$ having values in the commuting (bosonic) and anticommuting (fermionic) sector of an infinite dimensional Grassmann algebra.

The Kuper-KdV equation is:
\begin{equation}\label{k1}
u_t-6uu_x+u_{xxx}+12\xi\xi_{xx}=0
\end{equation}
\begin{equation}\label{k2}
\xi_t+4\xi_{xxx}-6u\xi_x-3u_x\xi=0
\end{equation}
with $u(x,t)$ bosonic function and $\xi(x,t)$ fermionic one.

The complete integrability was established by the existence of the Lax pair \cite{kupers}. Indeed if
$$L=\partial_x^2+u(x,t)+\xi(x,t)\partial_x^{-1}\xi(x,t)$$
then
$$L_t=[(L^{3/2})_{+},L]$$ is equivalent with Kuper-KdV. $L$ is a pseudo-differential super-operator and $L^{3/2}_{+}$ is the purely differential part of the formal power 3/2.

This is not the only one super-extension (non-supersymmetric). There is also the Holod-Pakuliak system \cite{holod}:
\begin{equation}\label{hp}
u_t= -u_{xxx} + 6uu_x + 6 ( \alpha \beta_{xx}-\alpha_{xx} \beta)
\end{equation}
$$\alpha_t =-4 \alpha_{xxx} + 6u\alpha_{x} + 3u_x \alpha$$
$$\beta_t = -4\beta_{xxx} + 6u\beta_x + 3u_x\beta$$
and the extended super-KdV of Geng-Wu \cite{geng}:
$$u_t = - u_{xxx} + 6uu_x + 12u \xi_{xx}\xi + 6u_x\xi_x \xi - 3 \xi_{xxxx}\xi - 6\xi_{xxx}\xi_x$$
$$\xi_t = - 4\xi_{xxx} + 3u_x\xi + 6u\xi_x.$$

\subsection{Bilinear form of Kuper-KdV}

Consider the following nonlinear substitutions $u=-2\partial_x^2\log F(x,t)$ and $\xi(x,t)=G(x,t)/F(x,t)$ where $G(x,t)$ is 
a grassmann odd (anticommuting) function and $F(x,t)$ is  a grassmann even (commuting) function.
The bilinear form will be:

\be\label{bil}
(D_tD_x+D_x^4)F\cdot F+6 D_x G\cdot G=0
\ee
$$(D_t+D_x^3)G\cdot F+3D_x K\cdot F=0$$
$$D_x^2 G\cdot F-K F=0,$$
where $K(x,t)$ is an auxiliary odd function. It has no role in the solution  but is crucial for bilinearisation.

{\bf Proof:}
Introducing $u=-2(\log F)_{xx}$ in (\ref{k1}) we get after one integration with respect to $x$:
$$2(\log F)_{xt}+2(\log F)_{xxxx}+3(2(\log F)_{xx})^2=12\xi\xi_x.$$
Using the definition of Hirota bilinear operator (\ref{1}) and (\ref{2}), this relation is transformed in:
$$\frac{(D_xD_t+D_{x}^4)F\cdot F}{F^2}=12\frac{G D_x G\cdot F}{F^3}.$$
But, because $G$ is grassmann odd we have $(12GD_xG\cdot F)/F^3=12GG_x/F^2=-
6D_x G\cdot G/F^2$. 
Accordingly we obtain the first bilinear equation:
$$(D_tD_x+D_x^4)F\cdot F+6 D_x G\cdot G=0.$$
Now we can write (\ref{k2}) in the following way:
$$\xi_t+4\xi_{xxx}-6u\xi_x-3u_x\xi=\xi_t+\xi_{xxx}-3u\xi_x+3(\xi_{xx}-u\xi)_x=0.$$
Using (\ref{4}) we get:
$$\frac{(D_t+D_{x}^3)G\cdot F}{F^2}+3(\xi_{xx}-u\xi)_x=0.$$
Let $K(x,t)$ be an auxiliary fermionic function such that:
$\xi_{xx}-u\xi=K/F$, which means (using (\ref{3})) $D_x^2 G\cdot F-KF=0$. Also, introducing above, we get:
$$(D_t+D_x^3)G\cdot F+3D_x K\cdot F=0.$$
Now we can compute directly the soliton solutions of the bilinear system (\ref{bil}) as combinations of exponentials $\exp(k_i x-\omega_i t)$, where $k_i$ are commuting (even) invertible grassmann numbers and $\omega_i=\omega_i(k_i)$ is 
the dispersion relation (some odd parameters can enter in the definition of the dispersion relation, but this is not the case here). 
For the fermionic (odd) tau functions $G(x,t)$ and $K(x,t)$, odd parameters $\zeta_i$ have to be considered. Accordingly, every super-soliton is characterized by the following triplet
$(k_i, \zeta_i, \omega_i(k_i,\zeta_i))$. Of course, nobody imposes the number of odd parameters for a soliton, but we consider the simplest case here where any soliton is characterized by only one $k$ and only one $\zeta$. 

The soliton solution has the following form:
\begin{itemize}
\item $1$-soliton solution 
$$G=\zeta_1 e^\frac{\eta_1}{2},\quad F=1+e^{\eta_1}, \quad K=\zeta_1 \frac{k_1^2}{4} e^\frac{\eta_1}{2}$$ where $\eta_1=k_1x-k_1^3 t$ and $\zeta_1$ is a free grassmann parameter
\item $2$-soliton solution
$$G=\zeta_1 e^\frac{\eta_1}{2}+\zeta_2 e^\frac{\eta_2}{2}+\zeta_1\alpha_{12}A_{12}e^{\frac{\eta_1}{2}+\eta_2}+\zeta_2 \alpha_{21}A_{21}e^{\frac{\eta_2}{2}+\eta_1}$$
$$F=1+e^{\eta_1}+e^{\eta_2}+A_{12}e^{\eta_1+\eta_2}+\zeta_1\zeta_2 A_{12}\frac{16}{(k_1-k_2)^3}e^\frac{\eta_1+\eta_2}{2}$$
$$K=\zeta_1 \frac{k_1^2}{4} e^\frac{\eta_1}{2}+\zeta_2 \frac{k_2^2}{4} e^\frac{\eta_2}{2}+\zeta_1 \frac{k_1^2}{4}\alpha_{12}A_{12}e^{\frac{\eta_1}{2}+\eta_2}+\zeta_2 \frac{k_2^2}{4}
\alpha_{21}A_{21}e^{\frac{\eta_2}{2}+\eta_1}$$
\item $3$-soliton solution
$$G=\sum_{i=1}^3 \zeta_i e^\frac{\eta_i}{2}+\sum_{i\neq j \neq l}^3\zeta_i \alpha_{ij}A_{ij}e^{\frac{\eta_i}{2}+\eta_j}(1+\alpha_{il}A_{il}A_{jl}
e^{\eta_l})+\zeta_1\zeta_2\zeta_3 M_{123}e^\frac{\eta_1+\eta_2+\eta_3}{2}$$
$$F=1+\sum_{i=1}^3 e^{\eta_i}+\sum_{i<j}^3A_{ij} e^{\eta_i+\eta_j}+A_{12}A_{13}A_{23}e^{\eta_1+\eta_2+\eta_3}+\sum_{i<j\neq l}\frac{16\zeta_i\zeta_j A_{ij}}{(k_i-k_j)^3}e^{\frac{\eta_i+\eta_j}{2}}\left(1+A_{il}A_{jl}e^{\eta_k}\right) $$
$$K=\sum_{i=1}^3 \zeta_i\frac{k_i^2}{4} e^\frac{\eta_i}{2}+\sum_{i\neq j \neq l}^3\zeta_i\frac{k_i^2}{4} \alpha_{ij}A_{ij}e^{\frac{\eta_i}{2}+\eta_j}(1+\alpha_{il}A_{il}A_{jl}
e^{\eta_l})+\zeta_1\zeta_2\zeta_3 M_{123}Q_{123}e^\frac{\eta_1+\eta_2+\eta_3}{2}$$
\end{itemize}
where:
$$A_{ij}=\left(\frac{k_i-k_j}{k_i+k_j}\right)^2,\quad \alpha_{ij}=\frac{k_i+k_j}{k_i-k_j}$$
$$M_{123}=-8\prod_{i\neq j \neq l}^3(k_i-k_l)\left(\frac{\alpha_{ij}A_{ij}}{(k_i-k_j)^2}+\frac{\alpha_{lj}A_{lj}}{(k_l-k_j)^2}\right)$$
$$Q_{123}=\frac{1}{3}\sum_{i\neq  j\neq l}^3\left(\frac{k_l^2}{4}+\frac{A_{il}A_{jl}\alpha_{il}\alpha_{jl}(k_i-k_j)^4}{4(k_j-k_l)^2A_{ij}A_{il}\alpha_{il}+4(k_i-k_l)^2A_{ij}A_{jl}\alpha_{jl})}\right).$$

It is instructive to see what is the difference between super-soliton dynamics of Kuper-KdV and the one of suspersymmetric KdV of Manin-Radul.

Indeed, the nonlinear form of supersymmetric KdV (susy-KdV) is \cite{mathieu}:
\begin{eqnarray}\label{skdv}
u_t+6uu_x+u_{xxx}-3\xi\xi_{xx}=0\\
\xi_{t}+3(\xi u)_{x}+\xi_{xxx}=0\nonumber
\end{eqnarray}
If $u=2\partial_{x}^2\log F, \xi=G/F$, then the supersymmetric bilinear form is \cite{carstea0,fane}:
$$(D_t+D_{x}^3)G\cdot F=0$$
$$(D_tD_x+D_{x}^4)F\cdot F=(D_t+D_x^3)G\cdot G$$
and the two super-soliton solution has the following form \cite{carstea1}:
$$F=1+e^{\eta_1}+e^{\eta_2}+A_{12}(1+\beta_{12}\zeta_1\zeta_2)e^{\eta_1+\eta_2}$$
$$G=\zeta_1e^{\eta_1}+\zeta_2 e^{\eta_2}+\zeta_1 A_{12}\alpha_{12}e^{\eta_1+\eta_2}+\zeta_2\alpha_{21}A_{21}e^{\eta_1+\eta_2},$$
where $\eta=k_ix-k_i^3 t, A_{ij}=(k_i-k_j)^2/(k_i+k_j)^2, \alpha_{ij}=(k_i+k_j)/(k_i-k_j), \beta_{ij}=2/(k_j-k_i)$.

One can see immediately the similarities between the forms of the two solutions (of kuper-KdV and susy-KdV). However the asymptotology of interaction is completely different.  
Indeed, in the reference system of soliton 1, ($\eta_1$ fixed and $k_1<k_2$) we obtain:
$$\lim_{t\to +\infty}F=1+e^{\eta_1}, \lim_{t\to+\infty}G=\zeta_1e^{\eta_1}$$
$$\lim_{t\to -\infty}F=1+A_{12}(1+2\beta_{12}\zeta_1\zeta_2)e^{\eta_1}, \lim_{t\to-\infty}G=\zeta_2+A_{12}(\zeta_1\alpha_{12}+\zeta_2\alpha_{21})e^{\eta_1}.$$
Accordingly the interaction is elastic for the bosonic component with a fermionic correction of the phase shift (given by $\beta_{12}\zeta_1\zeta_2$), but for the fermionic component the interaction is not elastic. 
Not only the amplitude is dressed by the $\alpha_{ij}$ but a creation of a fermionic background in the fermionic tau function appears given by $\zeta_2$ (however the conservation of energy is still preserved since the 
fermionic background can be gauged-away due to supersymmetry).

In the case of Kuper-KdV the 2-super-soliton solution has the following form:

$$G=\zeta_1 e^\frac{\eta_1}{2}+\zeta_2 e^\frac{\eta_2}{2}+\zeta_1\alpha_{12}A_{12}e^{\frac{\eta_1}{2}+\eta_2}+
\zeta_2\alpha_{21}A_{21}e^{\frac{\eta_2}{2}+\eta_1}$$
$$F=1+e^{\eta_1}+e^{\eta_2}+A_{12}e^{\eta_1+\eta_2}+\zeta_1\zeta_2 A_{12}\frac{16}{(k_1-k_2)^3}e^\frac{\eta_1+\eta_2}{2}.$$
Again, considering the reference frame of the first soliton (i.e. $\eta_1$ fixed) and $k_1<k_2$ we get:
$$\lim_{t\to +\infty}F=1+e^{\eta_1}, \lim_{t\to+\infty}G=\zeta_1e^{\eta_1}$$
$$\lim_{t\to -\infty}F=1+A_{12}e^{\eta_1}, \lim_{t\to-\infty}G=\zeta_2A_{12}\alpha_{12}e^{\eta_1/2}.$$
This is a simpler interaction. Asymptotically, the bosonic soliton does not feel at all the presence of the fermionic one. 
But the fermionic soliton has not only a phase shift, but also a changing of amplitude 
from $\zeta_1$ to $\zeta_2$. What is really interesting is the presence of the fermionic dressing $\alpha_{ij}=(k_i+k_j)/(k_i-k_j)$ which appears both in supersymmetric KdV and Kuper-KdV. Moreover this fermionic 
dressing seems to be universal in the sense that appears in all bilinear super-equations analyzed so far in literature. In the discrete setting 
the fermionic dressing appears as well in the form $\alpha_{ij}=(e^{k_i+k_j}-1)/(e^{k_i}-e^{k_j})$ \cite{fane}.

{\bf Remark:}
Using the same procedure we can find the bilinear for the Holod-Pakuliak system (\ref{hp}). Namely if $\alpha=G_1/F, \beta=G_2/F, u=-2(\log F)_{xx}$ then we get:
$$(D_tD_x+D_x^4)F\cdot F+6 D_x G_1\cdot G_2=0$$
$$(D_t+D_x^3)G_1\cdot F+3D_x K_1\cdot F=0$$
$$(D_t+D_x^3)G_2\cdot F+3D_x K_2\cdot F=0$$ 
$$D_x^2 G_1\cdot F-K_1 F=0$$
$$D_x^2 G_2\cdot F-K_2 F=0,$$
where $K_1, K_2$ are two auxiliary fermionic functions.  The computation of soliton solutions goes on the same way as in the case of the one component system, except of some phases in 
the exponentials. However the solutions 
turn out to be rather trivial, i.e. $G_1=\pm G_2$ reducing the system to the one component case. So maybe a more complicated ansatz involving more grassmann parameters is needed.

\section{``Spinning'' FPU-problem}

In this section we are going to give a possible physical model for Kuper-KdV equation. Consider a chain of Fermi-Pasta-Ulam nonlinear oscillators  
(defined through $u_n(t)$) having also some internal degrees of freedom expressed through {\it real} grassmann fields $\psi_n(t)$. 
We assume that in addition to nonlinear interaction we have also a phononic mediated 
interaction of nearest neighbors grassmann fields (similar to celebrated Su-Schriffer-Heeger model of polyacetylene dynamics, although in our model there is no hopping of electrons).
The hamiltonian has the form:
$$H=\sum_n\left(\frac{p_n^2}{2m}+\frac{1}{2}\alpha^2(u_{n+1}-u_n)^2+\frac{\beta}{3}(u_{n+1}-u_n)^3+i\gamma\psi_n\psi_{n+1}(1+\delta(u_{n+1}-u_n))\right).$$

The equations of motion are giving by the following formula \cite{berezin}:
$$\ddot u_n=-\partial H/\partial u_n\qquad i\dot\psi=\partial H/\partial \psi_n$$
$$\ddot u_n=\alpha^2 (u_{n+1}+u_{n-1}-2u_n)+\beta(u_{n+1}+u_{n-1}-2u_n)(u_{n+1}-u_{n-1})+i\gamma\delta\psi_n(\psi_{n+1}+\psi_{n-1})$$
$$\dot \psi_n=\gamma(\psi_{n+1}-\psi_{n-1})+\gamma\delta(u_{n+1}\psi_{n+1}+u_{n-1}\psi_{n-1})-\gamma\delta u_{n}(\psi_{n+1}+\psi_{n-1}).$$
The imaginary number $i$ in the hamiltonian and first equation is important from the physical point of view showing the real character of the equation (even though all the fields are real), 
because $(i\psi_n\psi_{n+1})^*=-i\psi_{n+1}^*\psi_n^*=i\psi_n\psi_{n+1}$. 

We want to compute the rigorous continuum limit through the method of multiple scales. Indeed in the linear regime we have two branches of dispersion related to the bosonic and fermionic field:
$$\omega_b^2-4\alpha^2\sin^2(k/2)=0 \quad {\rm and}\quad \omega_f+2\gamma\sin k=0.$$ 
In the continuum limit the wave length become large compared to $n$ so we can consider $k=k_0\epsilon$. Accordingly, $kn-\omega_ft=k_0\epsilon (n+2\gamma t)-\gamma k_0^3\epsilon^3 t/3+{\cal{O}}({\epsilon}^5)$ 
suggesting the following stretched variables:
$$x=\epsilon(n+2\gamma t), \qquad \tau=\frac{\gamma}{3}\epsilon^3 t.$$
Also, because of the nonlinear interaction the fields also will be modified to:
$$u_n(t)\to \epsilon^p \Phi(x,\tau),\qquad \psi_n(t)\to \epsilon^q \zeta(x,\tau),$$
where $p,q$ are numbers to be determined in asymptotic balance.
From the first equation we get:
$$\frac{\gamma^2}{9}\epsilon^{p+6}\Phi_{\tau\tau}+\frac{4\gamma^2}{3}\epsilon^{p+4}\Phi_{x\tau}+4\gamma^2\epsilon^{p+2}\Phi_{xx}=
\alpha^2\epsilon^{p+2}\Phi_{xx}+\frac{\alpha^2}{12}\epsilon^{p+4}\Phi_{xxxx}+$$
$$+2\beta\epsilon^{2p+3}\Phi_x\Phi_{xx}+i\gamma\delta \epsilon^{2q+2}\zeta\zeta_{xx},$$
while the second fermionic equation gives:
$$\frac{\gamma}{3}\epsilon^{q+3}\zeta_{\tau}+2\gamma\epsilon^{q+1}\zeta_x=\gamma\epsilon^q(2\epsilon \zeta_x+\frac{\epsilon^3}{3}\zeta_{xxx}+...)+$$
$$+\gamma\delta\epsilon^{p+q}(2\zeta\Phi+\epsilon^2(\zeta\Phi_{xx}+2\zeta_x\Phi_x+\zeta_{xx}\Phi)+...)-\gamma\delta\epsilon^{p+q}(2\zeta\Phi+\epsilon^2\zeta_{xx}\Phi+\frac{\epsilon^4}{12}\zeta_{xxxx}\Phi+...)$$
In order to have a nontrivial limit we have to cancel terms of order $\epsilon^{p+2}$ in the bosonic equation which means $\alpha^2=4\gamma^2$. Maximal balance principle \cite{martin} gives $p=1$ and $q=3/2$ and when we put 
$\epsilon\to 0$ we get:
$$\zeta_{\tau}=\zeta_{xxx}+\frac{\delta}{3}(u_x\zeta+2\zeta_x u)$$
$$\Phi_{x\tau}=\frac{1}{4}\Phi_{xxxx}+\frac{3\beta}{\gamma^2}\Phi_{x}\Phi_{xx}+\frac{3i\delta}{4\gamma}\zeta\zeta_{xx}.$$
Taking $\Phi_x=u(x,t), \tau\to 4\tau$ we get  the Kuper-KdV equation (up to definition of sign and parameters $\beta, \gamma, \delta$).

{\bf Remark:} Defining stretched variables using the other branch of the dispersion the final result is not changed. Also there is no way to find supersymmetric KdV (\ref{skdv}) from the asimptotology no matter how 
one chooses the parameters.

In addition, we have to point out that the above lattice hamiltonian is {\it not} modeling a phonon mediated interaction between nearest neighbors {\it spins}. In the pseudoclassical description developed by 
Berezin and Marinov \cite{berezin} the spin is represented as an {\it even noninvertible (nilpotent)} grassmann quantity (the vector product is not zero since the components anticommutes):
$${\vec S}_n=-\frac{i}{2}({\vec \psi}_n\times {\vec \psi}_n),$$ 
where the vector $\vec\psi_n$ has three gassmann odd components $(\phi_n,\chi_n, \sigma_n)$ and the hamiltonian is given by: 
$$H=\sum_n\left(\frac{p_n^2}{2m}+\frac{1}{2}\alpha^2(u_{n+1}-u_n)^2+\frac{\beta}{3}(u_{n+1}-u_n)^3+i\gamma({\vec S_n}\cdot{\vec S_{n+1}})(1+\delta(u_{n+1}-u_n))\right).$$

The equations of motion are:
$$\frac{d^2}{dt^2}u_n=\alpha^2 (u_{n+1}+u_{n-1}-2u_n)+\beta(u_{n+1}+u_{n-1}-2u_n)(u_{n+1}-u_{n-1})+i\gamma\delta{\vec S_n}\cdot({\vec S_{n-1}}-{\vec S_{n+1}})$$
$$\frac{d}{dt}{\vec S_n}={\vec S_n}\times({\vec S_{n+1}}+{\vec S_{n-1}})+\gamma\delta(u_{n+1}-u_n)({\vec S_n}\times{\vec S_{n+1}})+\gamma\delta(u_n-u_{n-1})({\vec S_n}\times{\vec S_{n-1}})$$
and in the continuum limit these equations goes to a coupled Boussinesq-Heisenberg ferromagnet system in grassmann algebra. Its integrability and solutions are open problems.

\appendix

\section{Hirota operator for superfunctions}
Hirota derivative is defined for superfunctions in the same way as for ordinary functions:
$$D_x^n a\cdot b=(\partial_x-\partial_y)^na(x)b(y)|_{x=y}$$ for any grassmann odd or grassmann even superfunctions $a(x), b(x)$
However unlike the ordinary case $D_x^{2n} a\cdot a=0, D_x^{2n+1}a\cdot a\neq 0$ if $a$ is grassmann odd. 

Also \cite{hirota}:
\begin{equation}\label{1}
2(\log b)_{xx}=\frac{D_x^2 b\cdot b}{b^2},\quad 2(\log b)_{xt}=\frac{D_xD_t b\cdot b}{b^2}
\end{equation}

\begin{equation}\label{2}
2(\log b)_{xxxx}=\frac{D_x^4 b\cdot b}{b^2}-3(\frac{D_x^2 b\cdot b}{b^2})^2
\end{equation}
\begin{equation}\label{3}
\partial_x(\frac{a}{b})=\frac{D_x a\cdot b}{b^2}, \quad \partial_x^2(\frac{a}{b})=\frac{D_x^2 a\cdot b}{b^2}-\frac{a}{b}\frac{D_x^2 b\cdot b}{b^2}, 
\end{equation}
\begin{equation}\label{4}
\partial_x^3(\frac{a}{b})=\frac{D_x^3 a\cdot b}{b^2}-3\frac{D_x a\cdot b}{b^2}\frac{D_x^2 b\cdot b}{b^2}
\end{equation}
are valid for any superfunction $a(x,t)$ and any {\it invertible grassmann even superfunction} $b(x,t)$

\end{document}